\documentclass[aps,prb,reprint,amsmath,amssymb,superscriptaddress,floatfix]{revtex4-2}
\usepackage{graphicx}
\usepackage{bm}
\usepackage{natbib}
\usepackage{url}
\usepackage{textcase}
\usepackage{amssymb}
\usepackage{amsmath}
\usepackage{physics}
\usepackage{appendix}
\usepackage{xcolor}
\usepackage[colorlinks=true,linkcolor=blue,urlcolor=blue,citecolor=blue,anchorcolor=blue]{hyperref}

\begin{document}

\title{One-dimensional extended Hubbard model coupled with an optical cavity}

\author{Taiga Nakamoto}
\email{taiganakamoto@g.ecc.u-tokyo.ac.jp}
\affiliation{Department of Physics, The University of Tokyo, Hongo, Tokyo 113-0033, Japan}
\author{Kazuaki Takasan}
\email{kazuaki.takasan@phys.s.u-tokyo.ac.jp}
\affiliation{Department of Physics, The University of Tokyo, Hongo, Tokyo 113-0033, Japan}
\author{Naoto Tsuji}
\email{tsuji@phys.s.u-tokyo.ac.jp}
\affiliation{Department of Physics, The University of Tokyo, Hongo, Tokyo 113-0033, Japan}
\affiliation{RIKEN Center for Emergent Matter Science (CEMS), Wako 351-0198, Japan}

\date{\today}

\begin{abstract}
We study the one-dimensional extended Hubbard model coupled with an optical cavity, which describes an interplay of the effect of vacuum fluctuation of light and the quantum phase transition between the charge- and spin-density-wave phases. The ground state and excitation spectrum of the model are calculated by numerically exact tensor-network methods.
We find that the photon number of the ground state is enhanced (suppressed) along the quantum phase transition line when the light-matter coupling is comparable to (much smaller than) the cavity frequency.
We also show that the exciton peak in the optical conductivity and single photon excitation peak in the photon spectrum exhibit the vacuum Rabi splitting at resonance due to the light-matter interaction.
This behavior is in contrast to the case without excitons, where the photon spectrum is merely broadened without splitting due to the lack of a sharp resonance.
\end{abstract}
\maketitle

\section{Introduction}

Engineering and controlling quantum materials with external light driving have been attracting much attention in the field of condensed matter physics \cite{Nasu_book, Giannetti_Ultrafast_2016, Oka2019, delaTorre_Colloquium_2021}.
For example, there have been observations of light-induced insulator-metal transitions \cite{Cavalleri_Femtosecond_2001, Iwai_Ultrafast_2003}, light-induced superconducting-like states in cuprates \cite{Fausti_LightInduced_2011}, and light-induced anomalous Hall effect in graphene \cite{McIver_Lightinduced_2020}.
In these phenomena, light can be considered as a classical electromagnetic field that drives the system into an interesting nonequilibrium state.
On the other hand, there exist situations where quantum aspects of light play an important role, namely in quantum optics and cavity quantum electrodynamics (QED).
In cavity QED, one can confine photons in a cavity to couple them strongly with atoms, leading to hybridization between the photon and atomic degrees of freedom.

Recently, the idea of cavity QED has been extended to quantum materials, where one can control quantum many-body systems with quantum light (i.e., cavity quantum materials \cite{Schlawin_Cavity_2022,Bloch_Strongly_2022,Garcia-Vidal_Manipulating_2021,Mivehvar_Cavity_2021,Lu_Cavity_2025}).
In these systems, we can take advantage of the vacuum fluctuation of light, which has been absent in classical coherent states of light, to change material properties.
Previously, there have been experimental reports on cavity-modified transport properties in organic semiconductors \cite{Orgiu_Conductivity_2015} and in two-dimensional electron gases \cite{Paravicini-Bagliani_Magnetotransport_2019}, the change of chemical reaction rates \cite{Hutchison_Modifying_2012,Thomas_GroundState_2016,Thomas_Tilting_2019}, and exciton-polariton condensation \cite{Deng_Condensation_2002,Deng_Excitonpolariton_2010}.
More recently, the change of superconducting critical temperatures \cite{Thomas_Exploring_2025}, control of the metal-insulator transition \cite{Jarc_Cavitymediated_2023}, enhancement of ferromagnetism \cite{Thomas_Large_2021}, and breakdown of topological protection of edge states \cite{Appugliese_Breakdown_2022} have been observed in cavity quantum materials.

Based on these developments, various possibilities of controlling quantum phases have been theoretically explored, including superconductivity \cite{Sentef_Cavity_2018,Schlawin_CavityMediated_2019,Curtis_Cavity_2019,Gao_Higgs_2021,Andolina_Amperean_2024} , magnetism \cite{Sentef_Quantum_2020,Roman-Roche_Photon_2021}, ferroelectricity \cite{Latini_Ferroelectric_2021,Ashida_Quantum_2020} and Kondo effect \cite{Kuo_Kondo_2023,Mochida_Cavityenhanced_2024}.
Moreover, it has been proposed that one can realize cavity-mediated long-range interactions \cite{Gao_Higgs_2021,Ciuti_Cavitymediated_2021,Chakraborty_LongRange_2021}, control of topological properties by chiral cavity \cite{Wang_Cavity_2019,Hubener_Engineering_2021,Masuki_Berry_2023}, and quantum Floquet engineering \cite{Sentef_Quantum_2020,Eckhardt_Quantum_2022} in the cavity quantum materials.
On the other hand, the relation between a quantum phase transition (such as those including charge-density-wave and spin-density-wave phases) and the photon vacuum fluctuation has not been well explored. This is of particular interest since various physical quantities of electron systems exhibit singular behavior near the transition, which may affect cavity photon properties, leading to a possibility of detecting a quantum phase transition by cavity photon measurements. 

In order to understand the interplay between cavity photons and quantum phase transitions, we study the one-dimensional extended Hubbard model coupled with an optical cavity (Fig.~\ref{fig:Cavity_setup}), using numerically exact tensor-network methods.
The one-dimensional extended Hubbard model is a prototypical model of strongly correlated electrons showing a quantum phase transition between charge-density-wave and spin-density-wave phases \cite{Nakamura_Tricritical_2000,Ejima_Phase_2007}. The model contains on-site and nearest-neighbor interactions, the latter of which helps to form excitons in excited states, as observed in the optical conductivity spectrum \cite{Essler_Excitons_2001,Jeckelmann_Optical_2003,Sugimoto_Pumpprobe_2023}.
We examine the behavior of cavity photons around the phase transition by studying the virtual photon number and the vacuum Rabi splitting.
The former counts the number of the ground state photons induced by the strong light-matter interaction.
The latter corresponds to the excitation energy splitting due to the energy exchange between the material quasiparticles and the cavity photons.
The strong electron-electron correlation effects in cavity quantum materials are studied mainly based on the Hubbard model, which only contains the on-site interaction \cite{Kiffner_Manipulating_2019,Kiffner_Mott_2019,Sentef_Quantum_2020,LeDe_Cavity_2022}.
However, the interplay between the cavity photons and quantum phase transitions as well as excitons, which do not appear in the one-dimensional half-filled Hubbard model, remains largely unexplored.

The rest of the paper is organized as follows.
In Sec.~\ref{sec:Model and Methods}, we introduce the low-energy effective model, in which the electrons with on-site and nearest-neighbor interactions are coupled to a single-mode photonic field.
We also briefly explain the numerical methods to calculate the ground state and the excited state of the light-matter coupled system.
In Sec.~\ref{sec:Virtual Photons}, we present the numerical results of the ground state that indicate the quantum phase transition has an effect on the virtual photon properties such as the photon number and the photon squeezing.
The Wigner function and the photon distribution function are also changed by the electronic interaction.
We also discuss the perturbation theory combined with the squeeze transformation to understand the behavior of the virtual photons.
In Sec.~\ref{sec:Vacuum Rabi Splitting}, we present the numerical results of the response function.
We find that there are two distinctive behaviors of the spectral function, vacuum Rabi splitting and spectral broadening, depending on the electronic interaction.
In Sec.~\ref{sec:Summary}, we summarize the results and discuss experimental relevance and future direction.

\section{Model and Methods} \label{sec:Model and Methods}

\begin{figure}[t]
  \begin{center}
  \includegraphics[width=0.483\textwidth]{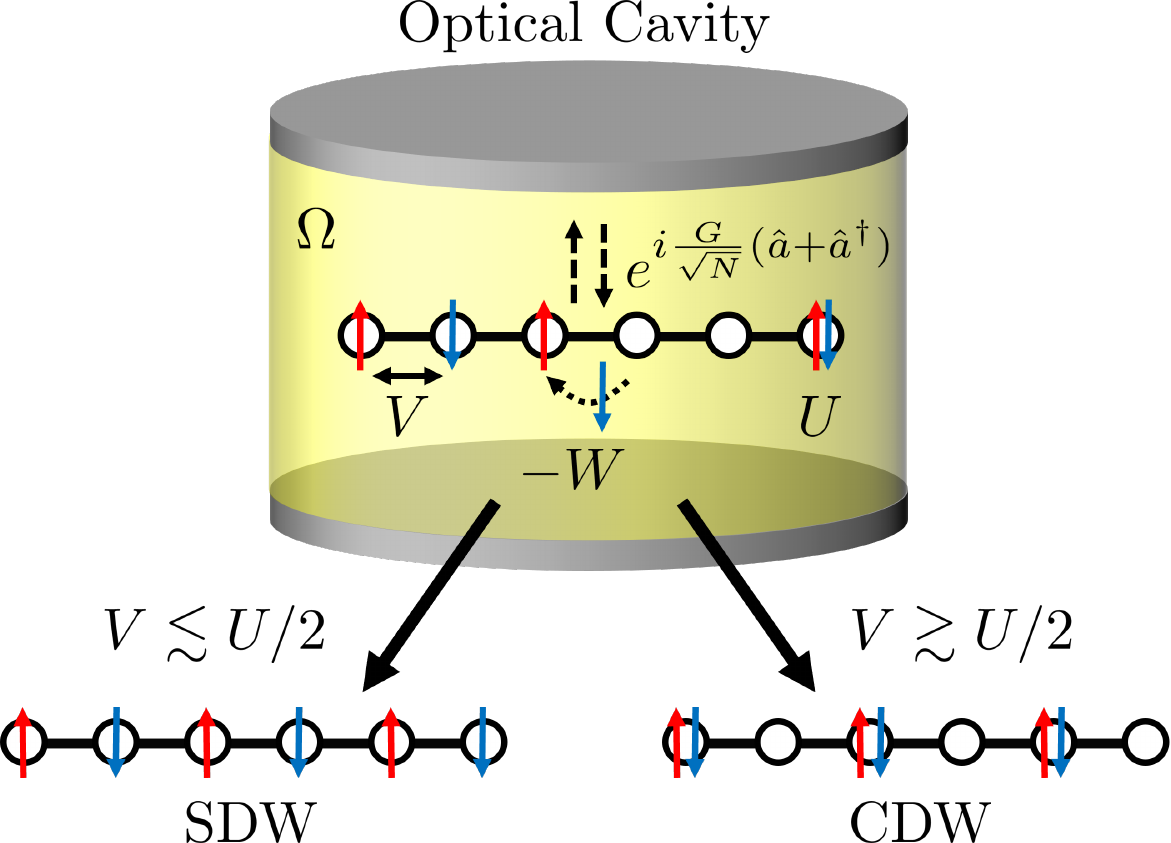}
  \caption{
  Schematic picture of the one-dimensional extended Hubbard system confined in a cavity, where the quantized electromagnetic field with frequency $\Omega$ is coupled to electrons with the coupling constant $G$.
  $W$, $U$, and $V$ are the hopping amplitude, on-site interaction, and nearest-neighbor interaction, respectively.
  The SDW phase appears for $V \lesssim U/2$, while the CDW phase appears for $V \gtrsim U/2$.} \label{fig:Cavity_setup}
  \end{center}
\end{figure}
We consider a model of interacting one-dimensional lattice electrons coupled with an optical cavity as illustrated in Fig. \ref{fig:Cavity_setup}.
Photons propagate between two mirrors in the cavity.
The energy of the cavity frequency is quantized as $\Omega(n,\vb{k}_{\parallel}) = c \sqrt{(\pi n/L_z)^2 + \vb{k}_{\parallel}^2}~(n=0,1,2,\cdots)$, where $c$ is the speed of light, $L_z$ is the length of the cavity, and $\vb{k}_{\parallel}$ is the wave vector parallel to the cavity mirror.
To minimally describe the degrees of freedom of the photons, we only consider the first transmittance of the cavity, $n=1$, and wave vector as $\vb{k}_{\parallel} = \vb{0}$ (long-wavelength approximation).
These approximations are justified when we are interested in the low-energy physics of the cavity-matter systems \cite{Rokaj_Light_2018,Eckhardt_Quantum_2022}.

There are several ways to incorporate the light-matter interaction in the model \cite{Li_Electromagnetic_2020,Dmytruk_Gauge_2021,Ashida_Cavity_2021}.
In this study, we use the quantized version of the Peierls phase in the Coulomb gauge, which maintains the gauge invariance of the lattice model.
The total Hamiltonian ($\hbar=1$) is 
\begin{eqnarray}
  \hat{H} =&& \Omega\hat{a}^{\dagger}\hat{a} -W\sum_{j=1,\sigma}^{N}(e^{i\frac{G}{\sqrt{L}}(\hat{a}+\hat{a}^{\dagger})}\hat{c}^{\dagger}_{j,\sigma}\hat{c}_{j+1,\sigma} + \mathrm{h.c.}) \nonumber \\ 
  &&+\hat{H}_{\mathrm{int}}. \label{eq:Hamiltonian}
\end{eqnarray}
Here, $\hat{a}~(\hat{a}^{\dagger})$ is the annihilation (creation) operator for photons with mode $n=1$ and $\vb{k_{\parallel}}=\vb{0}$, and $\hat{c}_{i,\sigma}~(\hat{c}^{\dagger}_{i,\sigma})$ is the annihilation (creation) operator for electrons with spin $\sigma$ at site $i$.
Each operator satisfies the commutation relation $[ \hat{a},\hat{a}^{\dagger}] = 1$ and the anticommutation relation $\{\hat{c}_{i,\sigma},\hat{c}^{\dagger}_{j,\sigma'}\} = \delta_{i,j}\delta_{\sigma,\sigma'}$, respectively.
$\Omega=c\pi/L_z$ is the cavity frequency, $G$ is the collective coupling constant between the cavity and electrons, $W$ is the nearest-neighbor hopping amplitude, $N$ is the total number of electrons, and $L$ is the number of lattice sites.
The coupling strength to the cavity is proportional to $1/\sqrt{V_m}$, where $V_m$ is the mode volume of the cavity and is proportional to the length of the one-dimensional system parallel to the mirrors.
Hence the light-matter coupling can be put as $G/\sqrt{L}$ with $G$ being independent of $L$.
Since the light-matter coupling is proportional to $1/\sqrt{L}$, the ground-state properties of the electron system remain unchanged in the thermodynamic limit (excitation properties may change).
$\hat{H}_{\mathrm{int}}$ is the electron-electron interaction term given by
\begin{eqnarray}
  \hat{H}_{\mathrm{int}}=U \sum_{j=1}^{N} \hat{n}_{j,\uparrow} \hat{n}_{j,\downarrow} + V\sum_{j=1}^{N} \hat{n}_{j} \hat{n}_{j+1},
\end{eqnarray}
where $n_{i,\sigma} = \hat{c}^{\dagger}_{i,\sigma}\hat{c}_{i,\sigma}$ is the electron number operator with spin $\sigma$ at site $i$, and $n_{i} = \hat{n}_{i,\uparrow}+\hat{n}_{i,\downarrow}$ is the total electron number operator at site $i$.
$U~(V)$ is the on-site (nearest-neighbor) electron interaction strength.
In this study, we use the open boundary condition, and the electron system is set at half filling (i.e., $N=L$). 
In that case, the light matter coupling constant $G/\sqrt{L}$ becomes $G/\sqrt{N}$.

If there is no light-matter interaction ($G=0$), the electron system is described by the one-dimensional extended Hubbard model, whose ground and excited states are well studied numerically \cite{Ejima_Phase_2007,Nakamura_Tricritical_2000,Jeckelmann_Optical_2003,Essler_Excitons_2001,Sugimoto_Pumpprobe_2023}.
The on-site repulsion $U$ favors the antiferromagnetic ordering, while the nearest-neighbor repulsion $V$ promotes the charge ordering.
Due to their competition, the ground state of this model exhibits the spin-density wave (SDW) phase for $V \lesssim U/2$ and the charge-density wave (CDW) phase for $V \gtrsim U/2$.
There is also the bond-order wave (BOW) phase between the two phases, which is characterized by the alternating strength of the expectation value of the local kinetic energy on the bond.

Excitation spectra, such as the optical conductivity, also reflect the properties of the one-dimensional extended Hubbard model.
For small nearest-neighbor interactions ($0 < V < 2W$), the optical conductivity shows a broad excitation spectrum corresponding to the excitation from the lower to upper Hubbard bands. 
In the middle interaction regime ($2W < V < U/2$), a sharp excitation peak emerges below the charge gap, which corresponds to the energy level of doublon-holon bound states (i.e., excitons).
When one goes to the large nearest-neighbor interaction regime ($U/2 < V$), there appears an excitation peak corresponding to the energy required to dissociate a doublon and a holon in the CDW state.

If the light-matter coupling is turned on ($G>0$), the model becomes hard to tackle because it simultaneously includes the electron-electron and the photon-electron correlation. 
In this study, we incorporate these correlations in a numerically exact way based on the tensor network method.
The ground and excited states of the system are calculated by the density-matrix renormalization group (DMRG) \cite{White_Density_1992,White_Densitymatrix_1993} and time-evolving block decimation (TEBD) methods \cite{Vidal_Efficient_2004}, which represent a wave function in terms of a matrix product state (MPS).
We set the photon degrees of freedom at one end of the MPS \cite{Passetti_Cavity_2023,Chiriaco_Critical_2022,Bacciconi_Firstorder_2023}.
In the TEBD simulation, we use a dynamical deformation of the MPS structure using swap gates \cite{Halati_Numerically_2020,Halati_Theoretical_2020,Halati_Breaking_2022,Bezvershenko_Dicke_2021}.
The physical and numerical errors can be controlled through a finite cutoff of the photon Hilbert space $N_{\mathrm{max}}$ and the bond dimension $\chi$.
We confirm that the results are converged with $N_{\mathrm{max}}=15$ and $\chi=1000$ in the DMRG calculation for $G=6.0$, and with $N_{\mathrm{max}}=2$ and $\chi=400$ in the TEBD calculation for $G=0.5$ (See Appendix \ref{Appendix:B} for the details).
In the numerical implementation, we use functions from the ITensor library \cite{Fishman_ITensor_2022}.

\section{Virtual Photons} \label{sec:Virtual Photons}

When the coupling constant is zero, the electrons and photons are decoupled, and there is no photon in the ground state.
As the coupling constant increases, there appear a finite number of photons even in the ground state, since the light-matter interaction modifies the vacuum of the system.
These ground-state photons are called virtual photons \cite{Ciuti_Quantum_2005,Ashhab_Qubitoscillator_2010,Stassi_Spontaneous_2013}.
The hybridization between electrons and photons will generate virtual photons so that the physical quantities of the ground-state photons are expected to reflect the properties of the electron system.
In order to confirm this, we calculate the number of photons, the photon squeezing, the Wigner function, and the photon distribution function for the extended Hubbard model coupled to a cavity with various Coulomb interaction parameters $U$ and $V$.

\subsection{Number of photons}
\begin{figure*}[t]
    \centering
    \includegraphics[width=0.95\textwidth]{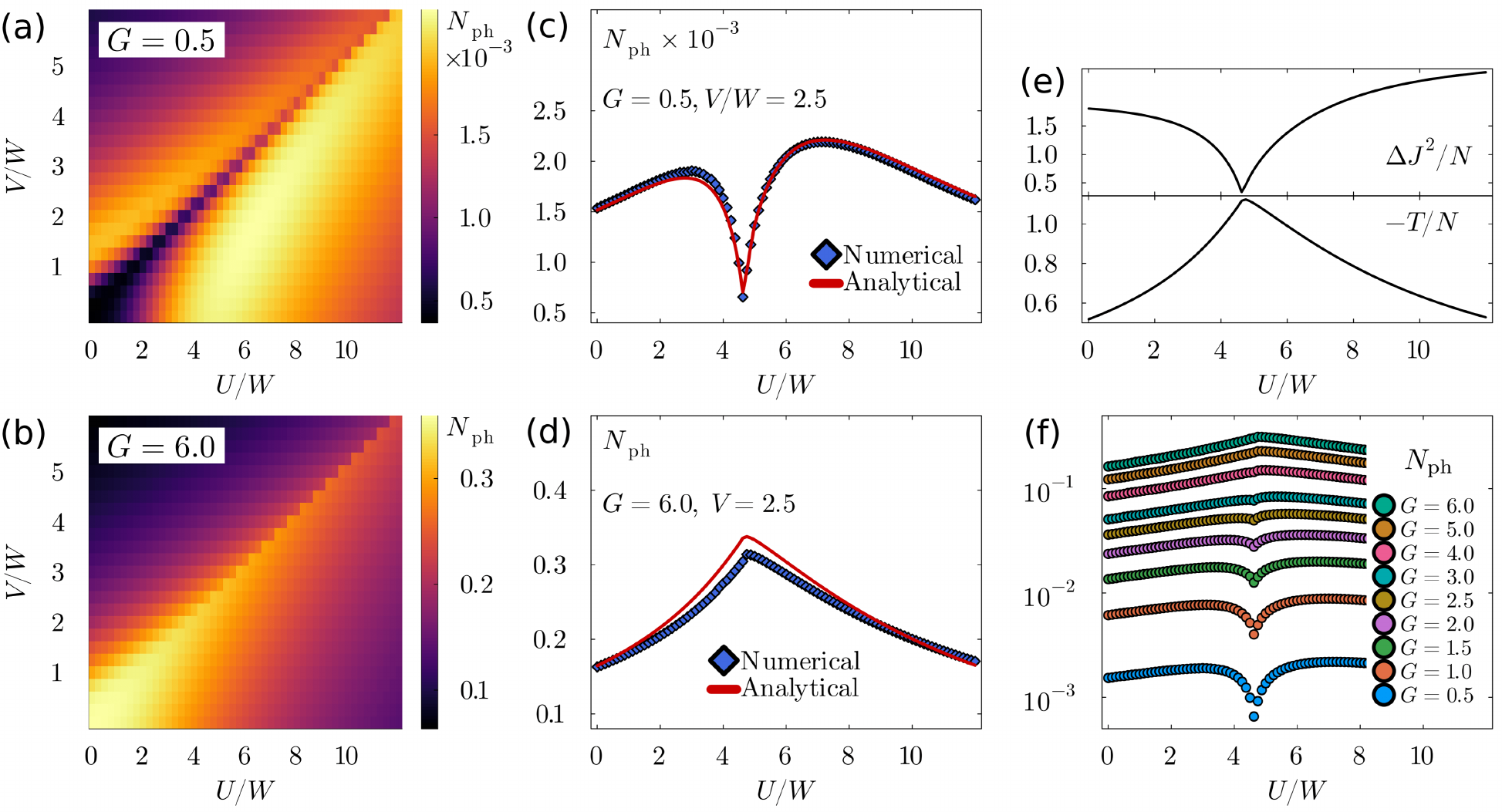}
    \caption{
    [(a), (b)] Photon number $N_{\rm ph}$ as a function of $U$ and $V$ for the one-dimensional extended Hubbard model coupled to an optical cavity with the coupling constant (a) $G=0.5$ and (b) $G=6$. 
    [(c), (d)] Comparison of the photon number calculated from the DMRG (blue marks) and the perturbation combined with the squeeze transformation (red lines) as a function of $U/W$ with $V/W=2.5$ for (c) $G=0.5$ and (d) $G=6$. 
    (e) The current fluctuation (top panel) and the kinetic energy (bottom panel) for the one-dimensional extended Hubbard model without a cavity.
    (f) Dependence of the photon number on the coupling constant $G$ plotted in the log scale.
    The results are obtained for the system size $N=80$ and the cavity frequency $\Omega/W = 10$.} \label{fig:Nph}
\end{figure*}
First, we numerically calculate the number of photons in the ground state, which is defined as
\begin{eqnarray}
    N_{\mathrm{ph}} &=& \left< \hat{a}^{\dagger}\hat{a} \right>_{\mathrm{GS}} \nonumber \\
    &=&\bra{\psi_{\mathrm{GS}}} \hat{a}^{\dagger}\hat{a} \ket{\psi_{\mathrm{GS}}},
\end{eqnarray}
where the ground-state wavefunction $\ket{\psi_{\mathrm{GS}}}$ is obtained by the DMRG method.
In Fig.~\ref{fig:Nph}, we show the photon number as a function of $U$ and $V$ for the coupling constant much smaller than the cavity frequency ($G^2 \ll \Omega/W$, shown in Fig.~\hyperref[fig:Nph]{2(a)}) and for the coupling constant comparable to $\Omega$ ($G^2 \gtrsim \Omega/W$, in Fig.~\hyperref[fig:Nph]{2(b)}), the latter of which corresponds to the ultrastrong coupling regime \cite{Forn-Diaz_Ultrastrong_2019, FriskKockum_Ultrastrong_2019, Qin_Quantum_2024} in experiments.
For $G^2 \ll \Omega/W$, the photon number is suppressed along the diagonal line ($V=U/2$), and takes large values slightly away from this line.
On the other hand, for $G^2 \gtrsim \Omega/W$, the photon number is enhanced along the line.
This line separates the SDW and CDW phases of the electron system, indicating that the virtual photons reflect the quantum phase transition of the electron system.

In order to understand the behavior of the photon number, we perform the perturbation theory in terms of the coupling constant $G$ combined with the squeeze transformation \cite{Ashida_Cavity_2021}, which allows us to perform the perturbative expansion even in the regime of $G^2 \gtrsim \Omega/W$ (see Appendix \ref{Appendix:A} for the details).
The photon number is expressed as
\begin{eqnarray}
  N_{\mathrm{ph}} &=& \sinh^2(\zeta) - 2\sinh(\zeta)\cosh(\zeta) \left< \hat{b}^{\dagger}\hat{b}^{\dagger}\right>_{\mathrm{GS}} \nonumber \\
  &&+ \qty(\cosh^2(\zeta) + \sinh^2(\zeta)) \left<\hat{b}^{\dagger}\hat{b}\right>_{\mathrm{GS}} \label{eq:Nph_sq},
\end{eqnarray}
where $\hat{b}$ ($\hat{b^{\dagger}}$) is the annihilation (creation) operator for squeezed photons defined as 
\begin{eqnarray}
  \mqty(\hat{b} \\ \hat{b}^{\dagger}) &=&\mqty(\cosh(\zeta) & \sinh(\zeta) \\ \sinh(\zeta) & \cosh(\zeta))\mqty(\hat{a} \\ \hat{a}^{\dagger}), \label{eq:def_squeezedop} \\
  \zeta &=& \frac{1}{2} \ln \qty(\widetilde{\Omega}/\Omega), \label{eq:zeta} \\
  \widetilde{\Omega} &=& \Omega \sqrt{1-2\frac{G^2T}{\Omega N}}. \label{eq:Omega_ren}
\end{eqnarray}
Here $\zeta$ is the squeeze factor, $\widetilde{\Omega}$ is the renormalized cavity frequency, and $T$ is the expectation value of the kinetic energy without the cavity as
\begin{eqnarray}
    \hat{T} &=& -W\sum_{j,\sigma}(\hat{c}^{\dagger}_{j,\sigma}\hat{c}_{j+1,\sigma}+\mathrm{h.c.}), \\
    T &=& \left< \hat{T} \right>_{\mathrm{GS},G=0}.
\end{eqnarray}
The expectation values of the squeezed photon number operator ($\hat b^\dagger \hat b$) and double creation operator ($\hat b^\dagger \hat b^\dagger$) are calculated by the second-order perturbation theory as
\begin{eqnarray}
  \left<\hat{b}^{\dagger}\hat{b}\right>_{\mathrm{GS}} &\simeq& \frac{\widetilde{G}^2}{N}\frac{\Delta J^2}{(\widetilde{\Omega}+\Delta E)^2}, \label{eq:bdagb} \\
  \left<\hat{b}^{\dagger}\hat{b}^{\dagger} \right>_{\mathrm{GS}} &\simeq& \frac{\widetilde{G}^2}{N}\frac{\Delta J^2}{\widetilde{\Omega}(\widetilde{\Omega}+\Delta E)} \label{eq:bdagbdag}, \\
  \Delta E &=& 
  \begin{cases}
    U-V & (V \leq U/2) \\
    3V-U & (V \geq U/2),
  \end{cases} \label{eq:ene_dif}
\end{eqnarray}
where $\widetilde{G} = G\sqrt{\Omega/\widetilde{\Omega}}$ is the renormalized coupling constant and $\Delta J^2$ is the expectation value of the current fluctuation without the cavity as
\begin{eqnarray}
    \hat{J} &=& -iW\sum_{j,\sigma}(\hat{c}^{\dagger}_{j,\sigma}\hat{c}_{j+1,\sigma}-\mathrm{h.c.}), \\
    \Delta J^2 &=& \left< \hat{J}^2 \right>_{\mathrm{GS},G=0} - \left< \hat{J} \right>_{\mathrm{GS},G=0}^2.
\end{eqnarray}
$\Delta E$ is the approximate charge excitation energy of the one-dimensional extended Hubbard model estimated in the large interaction limit ($U,~V \gg W$).

From Eqs.~\eqref{eq:Nph_sq}, \eqref{eq:zeta}, and \eqref{eq:Omega_ren}, we can understand the effect of the interaction on the photon number, ranging from $G^2 \ll \Omega/W$ to $G^2 \gtrsim \Omega/W$.
For $G^2 \ll \Omega/W$, there is almost no renormalization effect, i.e., $\widetilde{\Omega} \sim \Omega$ and $\widetilde{G} \sim G$.
Given that the squeeze factor $\zeta$ is small, $\sinh(\zeta) \sim 0$, and $\cosh(\zeta) \sim 1$, the leading contribution of the photon number is expressed as
\begin{eqnarray}
  N_{\mathrm{ph}} &\simeq& \frac{G^2}{N}\frac{\Delta J^2}{(\Omega+\Delta E)^2}. \label{eq:photon_smallG}
\end{eqnarray}
We compare the above analytical expression with the numerical DMRG results in Fig.~\hyperref[fig:Nph]{2(c)}.
The current fluctuation is suppressed around $V \simeq U/2$ between the SDW and CDW phases (Fig.~\hyperref[fig:Nph]{2(e)}, top panel) for the one-dimensional extended Hubbard model.
This is consistent with the suppression of the photon number at $V \simeq U/2$.
Due to the factor $\Delta E$ in the denominator of Eq.~\eqref{eq:photon_smallG}, $N_{\rm ph}$ is suppressed at both $U\rightarrow0$ and $U\rightarrow \infty$ since the charge excitation energy increases in these limits.
For $G^2 \gtrsim \Omega/W$, the renormalization effect is enhanced, and the leading contribution comes from the squeezed vacuum as
\begin{eqnarray}
  N_{\mathrm{ph}} &\simeq& \sinh^2(\zeta) \nonumber \\
  &\simeq& \frac{1}{4}\qty(\sqrt{1-2\frac{G^2T}{\Omega N}}-2). \label{eq:photon_largeG}
\end{eqnarray}
We compare the analytical expression with the numerical DMRG results in Fig.~\hyperref[fig:Nph]{2(d)}.
The half-filled electron system has the negative expectation value of the kinetic energy $T$, whose absolute value has large values around the BOW phase near $ V\simeq U/2$ (Fig.~\hyperref[fig:Nph]{2(e)}, bottom panel).
This is consistent with the enhancement of the photon number at $V \simeq U/2$.

Finally, we show crossover behavior from the regime of $G^2 \ll \Omega/W$ to the regime of $G^2 \gtrsim \Omega/W$ in Fig.~\hyperref[fig:Nph]{2(f)}.
The suppression of the photon number at the quantum phase transition line smoothly disappears around $G\sim 3.0$, and then turns into an enhancement.

\subsection{Photon squeezing}

As we see in the previous subsection, the photon squeezing is important for the coupling regime $G^2 \gtrsim \Omega/W$.
\begin{figure}[t]
  \centering
  \includegraphics[width=0.45\textwidth]{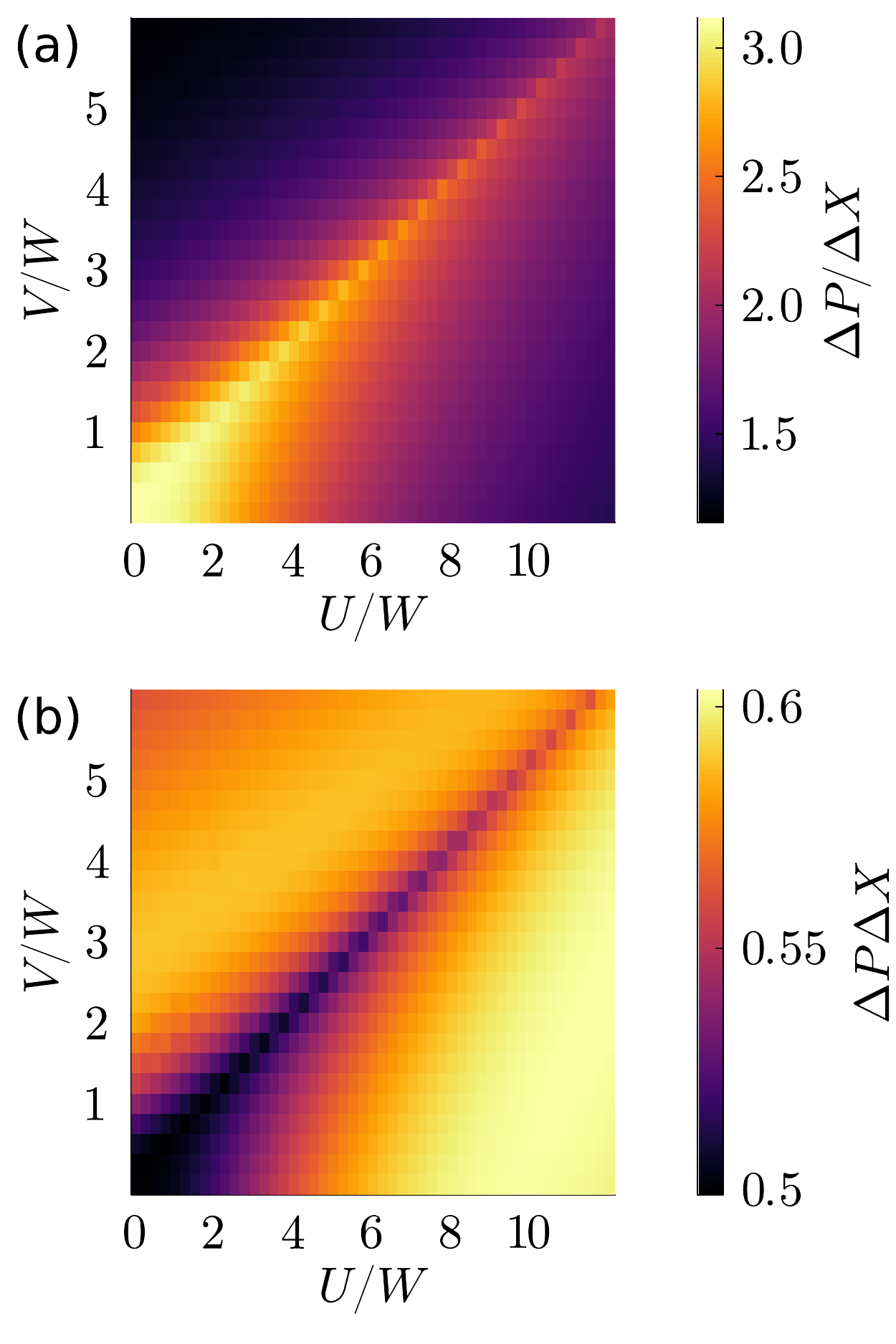}
  \caption{
  (a) $\Delta P/\Delta X$ and (b) $\Delta P\Delta X$ as a function of the on-site and nearest-neighbor interactions $U$ and $V$ for the one-dimensional extended Hubbard model coupled to an optical cavity with $N=80$, $\Omega/W=10$, and $G=6.0$.} \label{fig:photon_UVdep}
\end{figure}
In order to see the squeezing directly, we calculate the ratio between the variance of the canonical momentum and canonical coordinate operator, $\Delta P/\Delta X$, in Fig.~\hyperref[fig:photon_UVdep]{3(a)}.
The canonical momentum and coordinate operators are defined as $\hat{P}=i(\hat{a}-\hat{a}^{\dagger})/\sqrt{2}$ and $\quad \hat{X} = (\hat{a}+\hat{a}^{\dagger})/\sqrt{2}$, respectively.
In quantum optics, the product of $\Delta P$ and $\Delta X$ characterizes quantum fluctuations (as in the uncertainty relation).
Therefore, we also calculate $\Delta P \Delta X$ in Fig.~\hyperref[fig:photon_UVdep]{3(b)}.
$\Delta P / \Delta X$ is enhanced along the line $V=U/2$, which is similar to the electron interaction dependence of the photon number for $G^2 \gtrsim \Omega/W$ (Fig.~\hyperref[fig:Nph]{2(b)}).
On the other hand, $\Delta P \Delta X$ is suppressed along the diagonal line, which is similar to the electron interaction dependence of the photon number for $G^2 \ll \Omega/W$ (Fig.~\hyperref[fig:Nph]{2(a)}).

To understand the behavior of photon squeezing, we also perform the perturbation theory in terms of the coupling constant $G$ combined with the squeeze transformation.
Using the squeezed photon operator, $\Delta P / \Delta X$ and $\Delta P \Delta X$ are calculated as
\begin{eqnarray}
  \Delta P / \Delta X &=& e^{2\zeta}\sqrt{\frac{2\left<\hat{b}^{\dagger}\hat{b}\right>_{\mathrm{GS}} + 1-2\left<\hat{b}^{\dagger}\hat{b}^{\dagger}\right>_{\mathrm{GS}}}{2\left<\hat{b}^{\dagger}\hat{b}\right>_{\mathrm{GS}} + 1+2\left<\hat{b}^{\dagger}\hat{b}^{\dagger}\right>_{\mathrm{GS}}}} \nonumber \\
  &\simeq& \sqrt{1-2\frac{G^2T}{\Omega N}},
\end{eqnarray}
\begin{eqnarray}
  \Delta P \Delta X &=& \frac{1}{2} \sqrt{\qty(2\left<\hat{b}^{\dagger}\hat{b}\right>_{\mathrm{GS}} + 1)^2 - 4\left<\hat{b}^{\dagger}\hat{b}^{\dagger}\right>_{\mathrm{GS}}^2} \nonumber \\
  &\simeq& \frac{1}{2} + \frac{\widetilde{G}^2}{N}\frac{\Delta J^2}{(\widetilde{\Omega}+\Delta E)^2},
\end{eqnarray}
respectively.
From the perturbation results, we can understand the relation between $\Delta P / \Delta X$ (as well as $\Delta P \Delta X$) and $N_{\mathrm{ph}}$.
The ratio $\Delta P / \Delta X$ is enhanced along the quantum phase transition line due to the large absolute value of the kinetic energy.
The dependence on the parameters is similar to that of $N_{\rm ph}$ in Eq.~\eqref{eq:photon_largeG} except for the coefficient and constant shift.
The product $\Delta P \Delta X$ is suppressed along the quantum phase transition line due to the small value of the current fluctuation.
The dependence on the parameters is similar to that of $N_{\rm ph}$ in Eq.~\eqref{eq:photon_smallG} except for the presence of the standard quantum limit $1/2$.

\subsection{Wigner function and photon distribution function}

If there is no electron-electron interaction, the photon ground state is the squeezed vacuum state \cite{Eckhardt_Quantum_2022}.
However, once the electron interaction is turned on, the photon ground state includes a finite number of squeezed photons ($\left<\hat{b}^{\dagger}\hat{b}\right>_{\mathrm{GS}} \neq 0$) due to the presence of the nonzero current fluctuation (i.e., $\Delta J^2 \neq 0$) as in Eq.~\eqref{eq:bdagb}.
To describe the detailed behavior of the squeezed photons, we analyze the Wigner function and the photon distribution function.

The Wigner function is defined as
\begin{eqnarray}
  W(X,P) &=& \frac{1}{\pi} \int_{-\infty}^{\infty} \dd{y} e^{2i P y} \left< X+y \right| \hat{\rho}_{\mathrm{ph}} \left| X-y \right>,
\end{eqnarray}
where $\hat{\rho}_{\mathrm{ph}} := \Tr_e[\ket{\psi_{\mathrm{GS}}}\bra{\psi_{\mathrm{GS}}}]$ is the photon reduced density matrix obtained by tracing out the electron degrees of freedom.
$\left| x \right>$ is the eigenstate of the coordinate operator of photon $\hat{X}$.
The photon distribution function is defined as
\begin{eqnarray}
  P(n) = \Tr_{\mathrm{ph}}[\hat{\rho}_{\mathrm{ph}}\ket{n}\bra{n}],
\end{eqnarray}
where $\ket{n}$ is the $n$ photon Fock state.
For the squeezed vacuum state, the Wigner function and the photon distribution function are expressed as
\begin{eqnarray}
  W(X,P) &=& \frac{1}{\pi}\exp(-e^{2\zeta}X^2-e^{-2\zeta}P^2), \\
  P(n) &=&
  \begin{cases} 
    0 & (n ~:~ \mathrm{odd}) \\
    \frac{\tanh^{n}(\zeta)}{\cosh(\zeta)} \frac{(n)!}{2^{n}(\frac{n}{2})!^2} & (n ~:~ \mathrm{even}),
  \end{cases}
\end{eqnarray}
respectively.
\begin{figure}[t]
  \centering
  \includegraphics[width=0.483\textwidth]{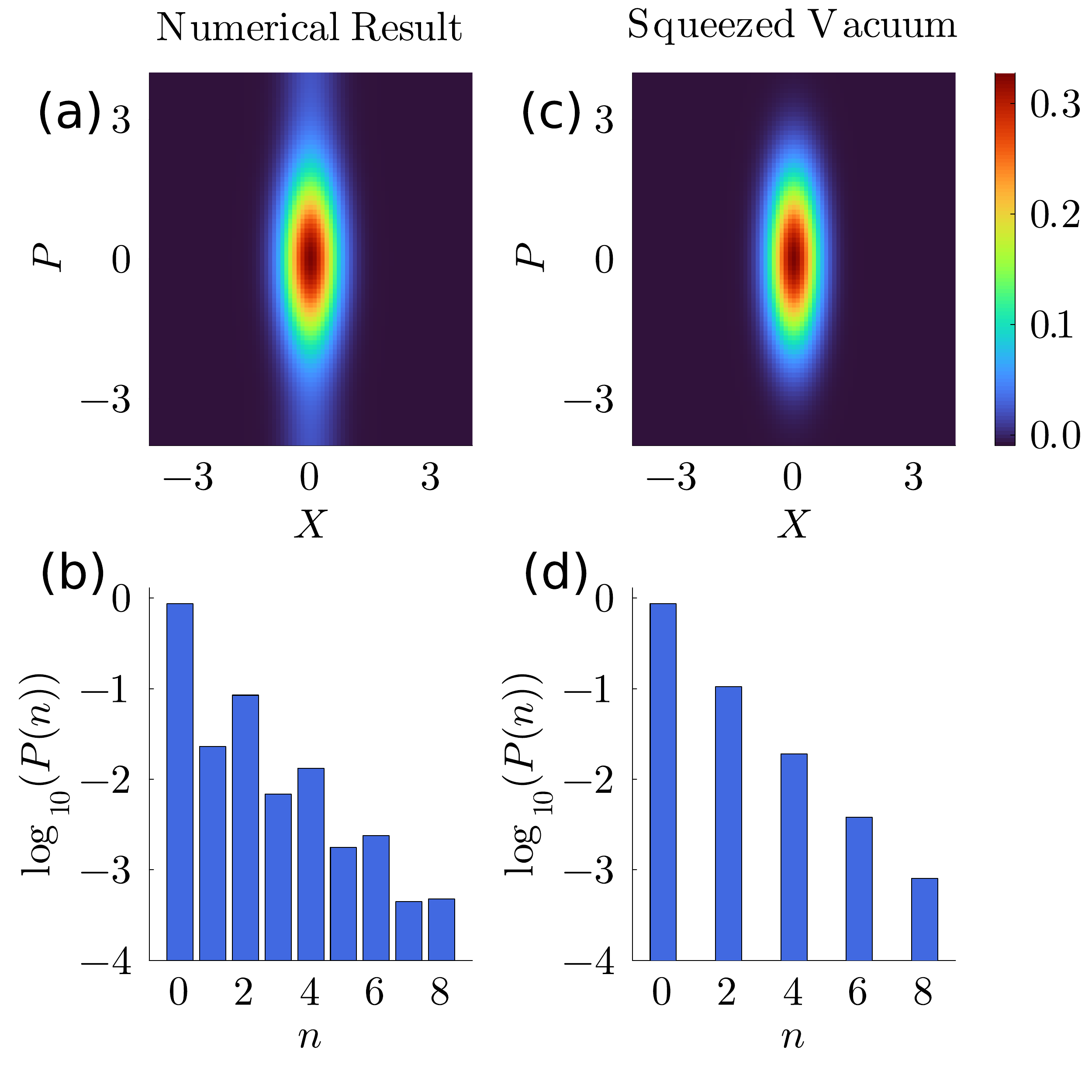}
  \caption{
  (a) Wigner function and (b) the photon distribution function calculated by DMRG for the one-dimensional extended Hubbard coupled to an optical cavity with $N=80$, $\Omega/W=10$, $G=6$, $U/W=10$, and $V/W=3$. 
  [(c), (d)] As a comparison, we show the corresponding results given by the squeezed vacuum state with the squeeze factor $\zeta$ [Eq.~\eqref{eq:zeta}].
  } \label{fig:wigPnph_UVdep}
\end{figure}
Here the Wigner function is given by the squeezed Gaussian function (Fig.~\hyperref[fig:wigPnph_UVdep]{4(c)}).
The photon distribution function has no odd-number photon states (Fig.~\hyperref[fig:wigPnph_UVdep]{4(d)}).

In Fig.~\ref{fig:wigPnph_UVdep}, we compare the Wigner function and the photon distribution function calculated from DMRG with those given by the squeezed vacuum with the same squeeze parameter $\zeta$.
Due to the electron interaction, the Wigner function is deformed from the squeezed Gaussian function, having higher canonical momentum contributions (Fig.~\hyperref[fig:wigPnph_UVdep]{4(a)}).
In spite of the high canonical momentum contribution, the total squeezing $\Delta P / \Delta X$ decreases from the case of vanishing electron interaction.
The photon distribution function includes contributions from odd-number photon states (Fig.~\hyperref[fig:wigPnph_UVdep]{4(b)}), whereas the squeezed vacuum state only includes even-number photon states.
If the electron interaction is present, the ground state changes from the Bloch state, and exhibits nonzero current fluctuation, which generates finite squeezed photon states including odd-number photon states.

\section{Vacuum Rabi Splitting} \label{sec:Vacuum Rabi Splitting}

Similar to the ground-state photonic properties studied in the previous section, the excitation spectra are modified due to the hybridization between electrons and photons.
Typically, the hybridization generates the vacuum Rabi splitting, which is the excitation energy splitting at the order of the coupling constant $G$.
In order to see that, we calculate the optical conductivity and the photon spectral function for the one-dimensional extended Hubbard model coupled with an optical cavity.
In these calculations, we only focus on the coupling constant much smaller than the cavity frequency ($G \ll \Omega/W$).

\subsection{Optical conductivity}

To define the optical conductivity, we consider a situation where a classical electromagnetic field is applied to the cavity-matter system.
The time-dependent Hamiltonian is given by replacing the Peierls phase as
\begin{eqnarray}
  i\frac{G}{\sqrt{L}}(\hat{a}+\hat{a}^{\dagger}) \longrightarrow i\frac{G}{\sqrt{L}}(\hat{a}+\hat{a}^{\dagger}) + iA(t),
\end{eqnarray}
where we adopt a weak vector potential $A(t)$ as follows,
\begin{eqnarray}
  A(t) = A_p\exp\qty(-\frac{(t-t_p)^2}{2\sigma_p^2})\cos\qty(\omega_p(t-t_p)).
\end{eqnarray}
A Gaussian-like envelope around $t=t_p$ is used with the temporal width $\sigma_p$ and the central frequency $\omega_p$.
We calculate the current $j(t)=\frac{1}{N}\left< \frac{\delta \hat{H}(t)}{\delta A(t)} \right>$ induced by the classical electromagnetic field.
Then, we perform the Fourier transformation of $j(t)$ and $A(t)$, and obtain the optical conductivity as
\begin{eqnarray}
  \sigma (\omega) &=& \frac{j(\omega)}{i(\omega+i\eta)A(\omega)}.
\end{eqnarray}
Note that a damping factor $e^{-\eta t}$ ($\eta$ is an infinitesimal positive constant) is introduced when the Fourier transformations are performed.
$\sigma(\omega)$ reflects the equilibrium property of the system, and does not depend on the details of the external field.

\begin{figure*}[t]
  \centering
  \includegraphics[width=0.95\textwidth]{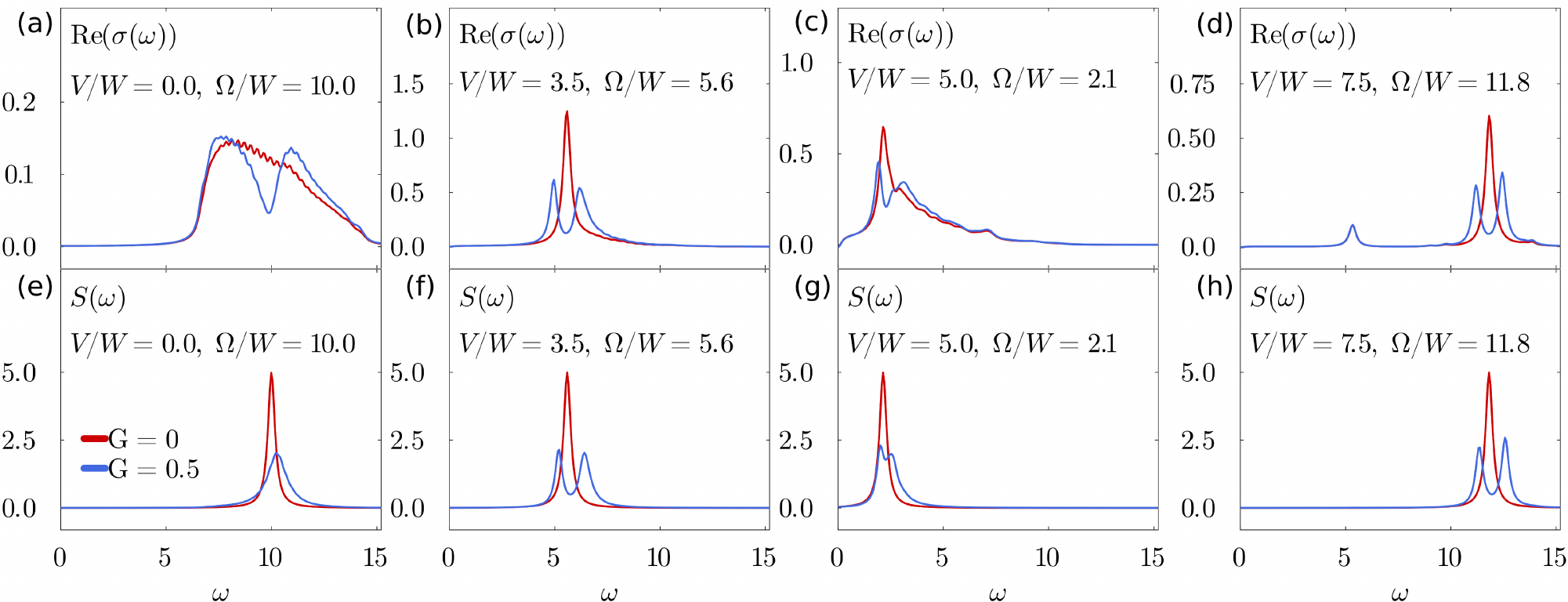}
  \caption{Real part of the optical conductivity $\mathrm{Re}(\sigma(\omega))$ ((a), (b), (c), (d)) and photon spectral function $S(\omega)$ ((e), (f), (g), (h)) for the one-dimensional extended Hubbard model coupled with the light-matter interaction (blue curves, $G=0.5$) with different nearest-neighbor interactions $V/W=0,~3.5,~5.0$, and $7.5$.
  The cavity frequencies are set to $\Omega/W=10,~5.6,~2.1$ and $11.8$, respectively, which are chosen from the peak center position of the optical conductivity without the light-matter interaction (red curves, $G=0$).
  The results are obtained for the system with $N=40$, $G=0.5$ and $U/W=10$, and the classical field parameters are $A_p=0.01,~t_p=1,~\sigma_p=0.02$, and $\omega_p=10$. The damping parameter is $\eta=0.2$} \label{fig:spectral}
\end{figure*}

In the Kubo formula, the optical conductivity (except for a possible singularity at $\omega=0$) is given as
\begin{eqnarray}
  \sigma(\omega) &=& \frac{i}{\omega N}\sum_{m} \Bigg(\frac{\abs{\bra{\psi_m}\hat{j}(0)\ket{\psi_\mathrm{GS}}}^2}{\omega+E_\mathrm{GS}-E_m+i\eta} \nonumber \\
  &&- \frac{\abs{\bra{\psi_m}\hat{j}(0)\ket{\psi_\mathrm{GS}}}^2}{\omega-E_{\mathrm{GS}}+E_m+i\eta} \Bigg),
  \label{eq:optical conductivity}
\end{eqnarray}
where $\ket{\psi_m}~(\ket{\psi_\mathrm{GS}})$ and $E_m~(E_\mathrm{GS})$ are the excited states (ground state) and their eigenenergies.
Therefore, the peak position of the optical conductivity represents the excited state energy induced by the current fluctuation.
We evaluate Eq.~\eqref{eq:optical conductivity} with the TEBD method.

We compare the real part of the optical conductivity for the one-dimensional extended Hubbard model with and without the light-matter interaction for $V=0$ in Fig.~\hyperref[fig:spectral]{5(a)}, $V=3.5W(<U/2)$
in Fig.~\hyperref[fig:spectral]{5(b)}, $V=5.0W(=U/2)$ in Fig.~\hyperref[fig:spectral]{5(c)}, and $V=7.5W(>U/2)$ in Fig.~\hyperref[fig:spectral]{5(d)}.
The blue curves show the results for the system with the light-matter interaction ($G=0.5$), while the red ones are the results without the light-matter interaction ($G=0$).
For $G=0$, the broad excitation at $V/W=0$ corresponds to the excitation energy from the lower to upper Hubbard bands.
The sharp excitation peak at $V/W=3.5$ corresponds to the energy level of excitons (i.e., doublon-holon bound states) in the SDW phase.
The frequency of the exciton peak takes the lowest value at the SDW-CDW transition point (near the BOW phase) around $V \simeq U/2$.
The sharp excitation peak at $V/W=7.5$ corresponds to the energy required to dissociate a doublon and a holon in the CDW phase.
The small peak around $\omega=5$ in Fig.~\hyperref[fig:spectral]{5(d)} is due to the finite-size effect. We have confirmed that the peak becomes smaller when we increase the system size.

We can see that the optical conductivity spectrum splits into two peaks with the spacing at the order of the coupling constant.
For $V/W=0$ (Fig.~\hyperref[fig:spectral]{5(a)}), the spectral splitting is consistent with the previous study \cite{Kiffner_Mott_2019}, which is based on the exact diagonalization method for small system sizes.
Using the TEBD method, we can see that the excitation continuum splits even for a relatively large system size.
We also find that sharper peak splitting occurs when the nearest-neighbor interaction is turned on.
For $V/W=3.5$ (Fig.~\hyperref[fig:spectral]{5(b)}), the peak splitting is similar to the case of the vacuum Rabi splitting of exciton-polaritons in semiconductors \cite{Deng_Excitonpolariton_2010}.
For $V/W=5.0$ (Fig.~\hyperref[fig:spectral]{5(c)}), the optical conductivity shows a complicated broad spectrum around the BOW phase, but we can still observe a splitting of the peak due to the coupling to the cavity.
For $V/W=7.5$ (Fig.~\hyperref[fig:spectral]{5(d)}), the peak splitting comes from the polariton composed of the photon and the excitation associated with the dissociation of the CDW state.

\subsection{Photon spectral function}

The photon spectral function is a measure of the absorptive response of the system against an external probe field.
The photon spectral function $S(\omega)$ is defined from the photon retarded Green's function $G^R(\omega)$ as
\begin{eqnarray}
  S(\omega) &=& -\mathrm{Im}(G^R(\omega)), \\
  G^R(t) &=& -i\theta(t)\bra{\psi_{\mathrm{GS}}}\comm{\hat{a}_{\mathrm{H}}(t)}{\hat{a}^{\dagger}_{\mathrm{H}}(0)}\ket{\psi_{\mathrm{GS}}},
\end{eqnarray} 
where $\hat{a}_{\mathrm{H}}(t)$ is the annihilation operator in the Heisenberg picture and $G^R(\omega)$ is the Fourier transform of $G^R(t)$.
In the Kubo formula, the photon retarded Green's function is given as
\begin{eqnarray}
  G^R(\omega) &=& \sum_{m}\Bigg(\frac{\abs{\bra{\psi_m}\hat{a}^{\dagger}\ket{\psi_\mathrm{GS}}}^2}{\omega+E_\mathrm{GS}-E_m+i\eta} \nonumber \\
  &&- \frac{\abs{\bra{\psi_m}\hat{a}\ket{\psi_\mathrm{GS}}}^2}{\omega-E_{\mathrm{GS}}+E_m+i\eta} \Bigg).
\end{eqnarray}
Therefore, the peak position of the photon spectral function represents the excited state energy induced by the photon creation and annihilation operators.

We compare the photon spectral function of the system with and without the light-matter interaction for $V=0$ in Fig.~\hyperref[fig:spectral]{5(e)}, $V=3.5W(<U/2)$ in Fig.~\hyperref[fig:spectral]{5(f)}, $V=5.0W(=U/2)$ in Fig.~\hyperref[fig:spectral]{5(g)}, and $V=7.5W(>U/2)$ in Fig.~\hyperref[fig:spectral]{5(h)}.
We can see that the photon spectral function splits when $V/W = 3.5$ and $V/W=7.5$.
In contrast, the photon spectral function does not show splitting for $V/W=0$ in spite of the splitting of the optical conductivity.
The photon spectral function becomes broadened, and its signal center is slightly shifted to the higher energy side.
For $V/W=5.0$, the splitting of the photon spectral function is weak but still visible.

In order to understand the splitting and broadening behavior, we analytically evaluate the retarded photon Green's function for a small coupli ng constant.
If the light-matter interaction is small, the main contribution to the spectral function comes from the electron's excited states with no photon $\ket{\phi_a}\otimes \ket{0}$ or the electron's ground state with a single photon $\ket{\phi_0}\otimes \ket{1}$.
The ground state would be given by the product state of the electron's ground state without photons $\ket{\phi_0}\otimes \ket{0}$ if we neglect virtual photons.
Then, we can write the photon retarded Green's function as
\begin{eqnarray}
  G^R(\omega) &=& \bra{0}\bra{\phi_0}\hat{a} \frac{1}{\omega-\hat{H}+E_{0}+i\eta} \hat{a}^{\dagger}\ket{\phi_0}\ket{0} \nonumber \\
  &=& \bra{1}\bra{\phi_0} \frac{1}{\omega-\hat{H}+E_{0}+i\eta} \ket{\phi_0}\ket{1} \label{eq:Green}.
\end{eqnarray}
In order to evaluate Eq.~\eqref{eq:Green}, we first consider the inverse of the matrix retarded Green's function in the bases of $\ket{\phi_0}\ket{1}$ and $\ket{\phi_a}\ket{0}~(a\neq0)$,
\begin{eqnarray}
  \hat{G}^{R,-1} &=& \mqty(z-\Omega & \frac{G_1}{\sqrt{N}} & \frac{G_2}{\sqrt{N}} & \cdots \\
                               \frac{G_1}{\sqrt{N}} & z- E_{10} &  0 & \cdots \\
                               \frac{G_2}{\sqrt{N}} & 0 &  z- E_{20} & \cdots \\
                               \vdots & \vdots & \vdots & \ddots), \label{eq:matrixGreen} \\ 
  G_a &=& \bra{\phi_a}\hat{J}\ket{\phi_0},
\end{eqnarray}
where $z = \omega + i\eta$, and $ E_{a0} = E_a-E_0$ is the energy difference between the excited states and ground state of the electron system.
In Eq.~\eqref{eq:matrixGreen}, we use the first-order expansion of the Peierls phase in terms of $G$.
By finding the (1,1) component of the inverse of Eq.~\eqref{eq:matrixGreen}, we can obtain the retarded Green's function [Eq.~\eqref{eq:Green}].
We use the cofactor expansion to find the (1,1) component of the inverse matrix, obtaining
\begin{eqnarray}
  G^R(\omega) &=& \frac{1}{\omega+i\eta-\Omega + \Sigma^R(\omega)}, \\
  \Sigma^R(\omega) &=& -\frac{1}{N}\sum_a\frac{\abs{G_a}^2}{\omega+i\eta- E_{a0}}.
\end{eqnarray}

\begin{figure}[t]
  \centering
  \includegraphics[width=0.483\textwidth]{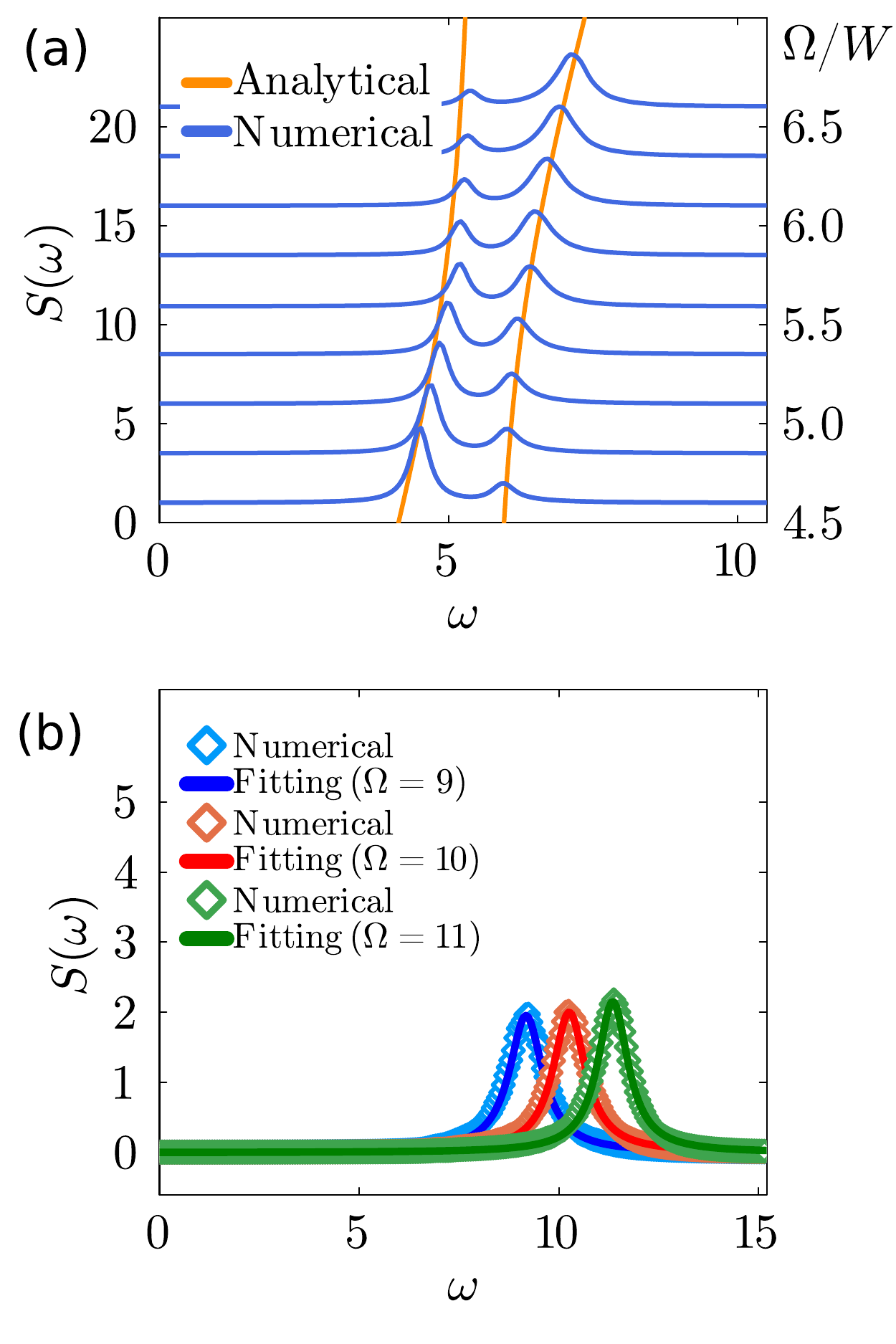}
  \caption{
  (a) Photon spectral function for the one-dimensional extended Hubbard model coupled to an optical cavity with various $\Omega/W$ (blue curves).
  The orange curves correspond to the position of the poles $E_{\pm}$ [Eq.~\eqref{eq:E_+-}]. We set $V/W=3.5$ and $E=5.6$. 
  (b) Photon spectral functions for various $\Omega/W$ (marks) are compared with the fitted Lorentzian function [Eq.~\eqref{eq:Lorentzian}] (solid curves).
  We set $V/W=0$ and take $N=40$, $G=0.5$, and $U/W=10$.}  \label{fig:spec_freqdep}
\end{figure}

If the excitation energy of the electron system induced by the current can be approximated as a single constant $E_{a0} \simeq E$ (as in the case of the red curves in Fig.~\hyperref[fig:spectral]{5(b)} and Fig.~\hyperref[fig:spectral]{5(d)}),
$\Sigma^{R}(\omega)$ becomes simplified as
\begin{eqnarray}
    \Sigma^R(\omega) &=& -\frac{G_{\mathrm{eff}}^2}{\omega+i\eta - E}, \\
    G_{\mathrm{eff}} &=& G\sqrt{<\Delta J^2>_{\mathrm{GS}}/N}.
\end{eqnarray}
$G_{\mathrm{eff}}$ is the effective coupling strength determined by the current fluctuation.
We find that the photon retarded Green's function has two poles $E_{\pm}$,
\begin{eqnarray}
    G^R(\omega) &=& \frac{\omega + i\eta - E}{(\omega+i\eta-E_+)(\omega+i\eta-E_-)}, \\
    E_{\pm} &=& \frac{\Omega+ E}{2} \pm \sqrt{\Delta^2 +  G_{\mathrm{eff}}^2}, \label{eq:E_+-} \\
    \Delta &=& \frac{\Omega - E}{2}.
\end{eqnarray}
The position of the two poles $E_{\pm}$ is nothing but the polariton energy of the Hopfield model \cite{Hopfield_Theory_1958}.
Compared to the Hopfield model, the splitting width depends on the current fluctuation of the electron system through the effective coupling strength. 
We compare the numerical TEBD results and $E_{\pm}$ for various cavity frequencies in Fig.~\hyperref[fig:spec_freqdep]{6(a)}.
We can see that the numerical calculations and the position of the two poles qualitatively agree with each other.

If the excitation energy $E_{a0}$ is distributed in a wide range of the frequency space (as in the case of the red curve in Fig.~\hyperref[fig:spectral]{5(a)}), the $\omega$ dependence of $\Sigma^R(\omega)$ can be neglected as $\Sigma^R(\omega) \sim \Sigma^R_0 + i \Sigma^R_1$, where $\Sigma^R_0$ and $\Sigma^R_1$ are the real and imaginary parts of $\Sigma^R(\omega)$.
We find that the photon spectral function becomes the Lorentzian, i.e.,
\begin{eqnarray}
  S(\omega) &=& \frac{\Sigma^R_1}{(\omega-\Omega+\Sigma^R_0)^2+(\Sigma^{R}_1)^2}. \label{eq:Lorentzian}
\end{eqnarray}
Figure \hyperref[fig:spec_freqdep]{6(b)} shows the result of the fitting of the spectral data obtained by the TEBD with a Lorentzian.
The Lorentzian well reproduces the spectral broadening and the shift of the spectral center, where $\Sigma_0^R$ corresponds to the shift of the spectral center and $\Sigma_1^R$ corresponds to the broadening of the spectral width.

\section{Summary and outlook} \label{sec:Summary}

To summarize, we have investigated the impact of the electron interaction on the light-matter hybridization through the properties of the virtual photons and vacuum Rabi splitting.
We consider the one-dimensional extended Hubbard model coupled to the cavity, and calculate the virtual photon number, photon squeezing, Wigner function, photon distribution function, optical conductivity, and photon spectral function.
Using the tensor-network method, we have obtained numerically exact results for those quantities.
At $U=V/2$, the current fluctuation is suppressed and the kinetic energy is enhanced, which leads to the suppression (enhancement) of the virtual photon number for $G^2 \ll \Omega/W$ ($G^2 \gtrsim \Omega/W$).
The photon squeezing ($\Delta P/\Delta X$) and quantum fluctuation ($\Delta P \Delta X$) have similar behavior to the virtual photon number for the same reason.
As a result, the photon number and the photon squeezing reflect the quantum phase transition in the one-dimensional extended Hubbard model.
We also find the differences between the Wigner function and the photon distribution function of the squeezed vacuum state and those of the numerical calculation.
The differences are caused by the electron interactions.

In the latter part of the paper, we study the vacuum Rabi splitting in the optical conductivity and the photon spectral function.
We find that the vacuum Rabi splitting always occurs in the optical conductivity, but the photon spectral function shows the splitting only when the nearest-neighbor interaction is turned on.
When the nearest-neighbor interaction is zero, the spectral function becomes broadened.
We have analytically obtained the retarded photon Green's function for a small coupling constant, which explains the splitting behavior and the broadening behavior consistently.

Our results demonstrate that the quantum phase transition from the spin density wave (SDW) to the charge density wave (CDW) phase is sharply reflected in the properties of cavity photons coupled to the electron system. 
This suggests a potential application for detecting the material phase transition using the optical cavity. 
Furthermore, it may lead to the control of quantum light via material phase transitions.
We also demonstrate that the exciton-polariton signal appears in the strongly correlated system as in semiconductor systems.
Our results will serve as a step toward exploring the exciton-polariton physics (such as their condensation \cite{Deng_Condensation_2002,Deng_Excitonpolariton_2010} and generation of lasers \cite{Schneider_Electrically_2013,Fraser_Physics_2016}) in strongly correlated materials.

Finally, we comment on the experimental relevance of our results.
The strong light-matter coupling has been observed \cite{Forn-Diaz_Ultrastrong_2019} through the Rabi splitting.
Organic salt \cite{Hasegawa_Electronic_1997,Hasegawa_Mixedstack_2000} and halogen-bridged transition-metal compounds \cite{Ono_Direct_2005,Iwai_Ultrafast_2003} have been considered to realize the one-dimensional extended Hubbard model.
Therefore, the observation of the vacuum Rabi splitting in these materials will be the first step to test the validity of our theoretical results.
In addition, it has recently become possible to directly observe the vacuum fluctuations inside a cavity \cite{Riek_Direct_2015,Benea-Chelmus_Electric_2019}.
Experimental tests of our predictions about the virtual photon are possible by observing the vacuum fluctuations in these materials.

\section{Acknowledgements}
We are grateful to Y. Ashida, M. Bamba, S. Imai, H. Katsura, K. Masuki, R. Matsunaga, J. Mochida, T. Morimoto, R. Nagashima, R. Ueda, and H. Ushihara for helpful discussions.
T. N. is supported by WINGS-MERIT of the University of Tokyo and JST SPRING (Grant No.~JPMJSP2108).
This work is supported by JST FOREST (Grant No.~JPMJFR2131), JST PRESTO (Grant No.~JPMJPR2256) and JSPS KAKENHI (Grant Nos.~JP22K20350, JP23K17664, JP24H00191, JP25H01246, JP25H01251, and JP25K17312).

\appendix

\section{Perturbation Theory} \label{Appendix:A}

In this Appendix, we show how to obtain the perturbation results combined with the squeeze transformation used in Sec. \ref{sec:Virtual Photons}.
We expand the Hamiltonian \eqref{eq:Hamiltonian} up to the quadratic order in $G$.
As long as we are interested in the collective coupling phenomena ($G \ll \sqrt{N}$), this expansion is justified in the thermodynamic limit $N \rightarrow \infty$ \cite{Eckhardt_Quantum_2022}.
Then, the total Hamiltonian is divided into the unperturbed part $\hat{H}_0$ and the perturbed part $\hat{H}_{\mathrm{pert}}$ as
\begin{eqnarray}
  \hat{H}_0 &=& \Omega\hat{a}^{\dagger}\hat{a} -\frac{G^2}{2N}T(\hat{a}+\hat{a}^{\dagger})^2 + \hat{H}_{\mathrm{e}}, \\
  \hat{H}_{\mathrm{pert}} &=& \frac{G}{\sqrt{N}}\hat{J}(\hat{a}+\hat{a}^{\dagger}) - \frac{G^2}{2N}\Delta \hat{T}(\hat{a}+\hat{a}^{\dagger})^2,
\end{eqnarray}
where $\hat{H}_{\mathrm{e}}$ is the Hamiltonian of the one-dimensional extended Hubbard model without cavity, $\hat{J} = -iW\sum_{j,\sigma}(\hat{c}^{\dagger}_{j,\sigma}\hat{c}_{j+1,\sigma}-\mathrm{h.c.})$ is the total current operator, $\hat{T} = - W \sum_{j=1}^{L}(\hat{c}^{\dagger}_{j,\sigma}\hat{c}_{j,\sigma}+\mathrm{h.c.})$ is the total kinetic energy operator, $T$ is the expectation value of the kinetic energy of the electron system without cavity, and $\Delta \hat{T} = \hat{T} - T$ is the fluctuation operator of the kinetic energy.
The diamagnetic contribution of the light-matter interaction is partially included in the unperturbed Hamiltonian.

The unperturbed Hamiltonian is decoupled into the photon and electron parts.
The squeeze transformation diagonalizes the former part as
\begin{eqnarray}
  \Omega\hat{a}^{\dagger}\hat{a} &-& \frac{G^2}{2N}T(\hat{a}+\hat{a}^{\dagger})^2 = \widetilde{\Omega}\hat{b}^{\dagger}\hat{b},
\end{eqnarray}
where $\hat{b}~(\hat{b}^{\dagger})$ is the annihilation (creation) operator of the squeezed photon defined in Eq.~\eqref{eq:def_squeezedop} in the main text, and
$\widetilde{\Omega}$ is the renormalized cavity frequency.
We note that the squeezed photon operator also satisfies the commutation relation, $[\hat{b},\hat{b}^{\dagger}]=1$.
Under this transformation, the unperturbed and perturbed Hamiltonians are rewritten in terms of the squeezed photon operator as
\begin{eqnarray}
  \hat{H}_0 &\rightarrow& \widetilde{\Omega}\hat{b}^{\dagger}\hat{b} + \hat{H}_{\mathrm{e}}, \\
  \hat{H}_{\mathrm{pert}} &\rightarrow& \frac{\widetilde{G}}{\sqrt{N}}\hat{J}(\hat{b}+\hat{b}^{\dagger}) - \frac{\widetilde{G}^2}{2N}\Delta \hat{T}(\hat{b}+\hat{b}^{\dagger})^2,
\end{eqnarray}
where $\widetilde{G} = G \sqrt{\Omega/\widetilde{\Omega}}$ is the renormalized cavity coupling constant.

This squeeze transformation has an advantage in performing the perturbation.
\begin{figure}[t]
\begin{center}
\includegraphics[width=0.45\textwidth]{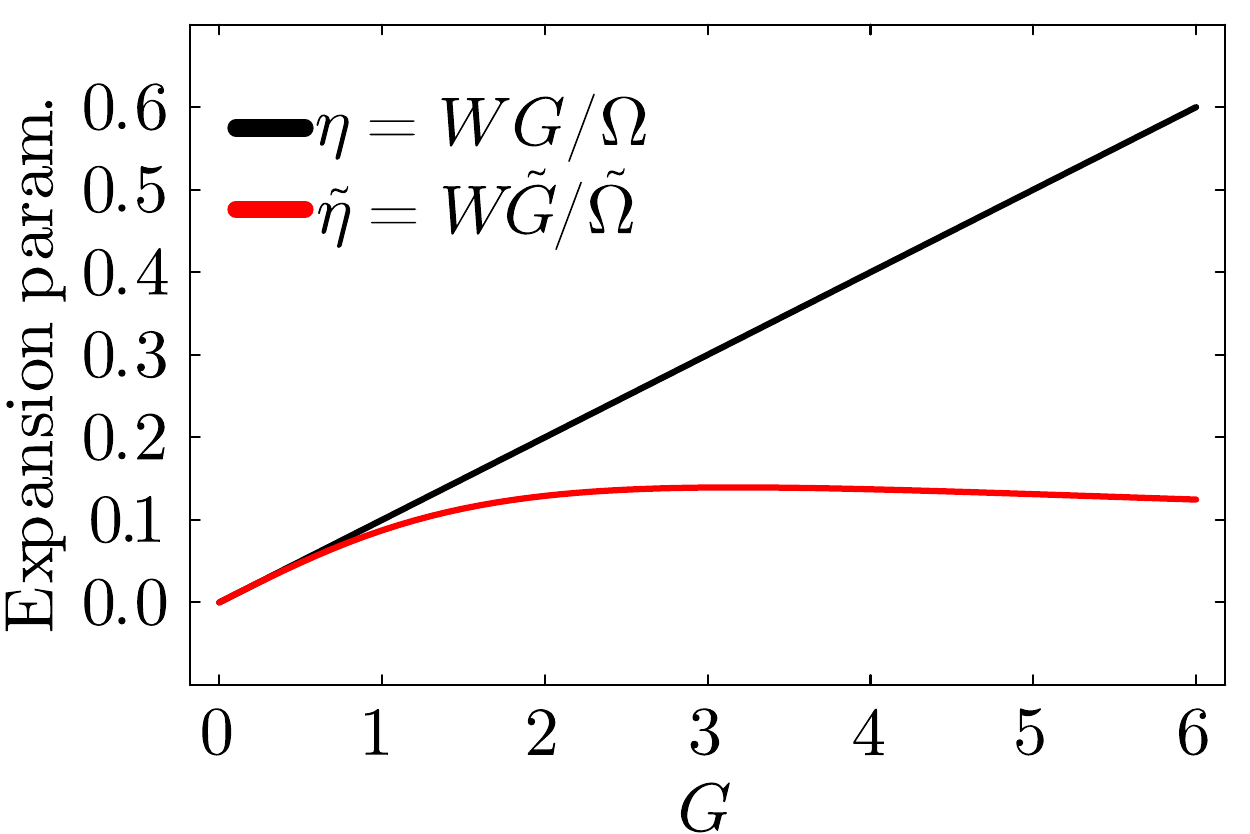}
\caption{
Expansion parameter $\eta=WG/\Omega$ and the renormalized expansion parameter $\widetilde{\eta} = W\widetilde{G}/\widetilde{\Omega}$ as a function of the coupling constant $G$.} \label{fig:perturbation_param}
\end{center}
\end{figure}
Figure \ref{fig:perturbation_param} shows the bare expansion parameter $\eta = WG/\Omega$ and the renormalized expansion parameter $\widetilde{\eta} = W \widetilde{G}/\widetilde{\Omega}$ as a function of the coupling constant $G$.
Compared to $\eta$, $\widetilde{\eta}$ is suppressed even for $G^2 \gtrsim \Omega/W$.
Therefore, the renormalized parameters allow us to perform the perturbation theory in all the range of the coupling constant.
We comment that a similar procedure has been employed previously in other light-matter coupled systems \cite{Ashida_Cavity_2021}.

The eigenstates of the unperturbed system are given by the product states of the electrons and the squeezed photon as $\ket{\psi_n^{(0)}} = \ket{\phi_a}\ket{l}$, where $\ket{\phi_a}$ is the eigenstate of the electrons with eigen energy $E_a$ and $\ket{l}$ is the Fock state of the squeezed photons with photon number $l$.
The corresponding eigenenergies are $E_n^{(0)}=E_a+l\widetilde{\Omega}$.
The second-order perturbation leads to the following squeezed photon number operator,
\begin{eqnarray}
    \left< \hat{b}^{\dagger}\hat{b} \right> &=& \sum_{n,m\neq0} \frac{ (H_{\mathrm{pert}})_{0,m}(H_{\mathrm{pert}})_{n,0} }{(E_0^{(0)}-E_n^{(0)})(E_0^{(0)}-E_m^{(0)})} \nonumber \\
    &&\times \bra{\psi_m^{(0)}} \hat{b}^{\dagger}\hat{b} \ket{\psi_n^{(0)}} \nonumber \\
    &\simeq& \frac{\widetilde{G}^2}{N}\sum_{a\neq0}\frac{\abs{\bra{\phi_a}\hat{J}\ket{\phi_0}}^2\abs{\bra{1}\hat{b}^{\dagger}\ket{0}}^2}{(\widetilde{\Omega}+ E_a-E_0)^2}.
\end{eqnarray}

Now, we make an approximation that the charge excitation energies are represented by a single energy $E_a-E_0 \rightarrow \Delta E$ as in Eq.~\eqref{eq:ene_dif} in the main text.
In the large Coulomb interaction limit ($U,~V \gg W$), the SDW state consists of singly occupied states.
Thus, the energy of the first-excited state, characterized by the presence of an adjacent doublon and holon, is approximately given by $U-V$.
On the other hand, doublons and holons align alternately in the CDW state.
Therefore, the first-excited state, where two adjacent sites become singly occupied, requires an excitation energy of approximately $3V-U$.
The sum over the excited state becomes
\begin{eqnarray}
    \sum_{a\neq0}\abs{\bra{\phi_a}\hat{J}\ket{\phi_0}}^2 &=& \bra{\phi_0}\hat{J^2}\ket{\phi_0} - \bra{\phi_0} \hat{J} \ket{\phi_0}^2 \nonumber \\
    &=& \Delta J^2. \label{eq:cur_fluc}
\end{eqnarray}
Then, we can obtain Eq.~\eqref{eq:bdagb} in the main text.
The expectation value of the double creation operator can also be calculated as
\begin{eqnarray}
    \left< \hat{b}^{\dagger}\hat{b}^{\dagger} \right> &=& \sum_{n,m\neq0} \frac{(H_{\mathrm{pert}})_{0,m}(H_{\mathrm{pert}})_{m,n}}{(E^{(0)}_0-E^{(0)}_m)(E^{(0)}_0-E^{(0)}_n)} \nonumber \\
    && \times \bra{\psi^{(0)}_n} \hat{b}^{\dagger}\hat{b}^{\dagger}\ket{\psi^{(0)}_0} \nonumber \\
    &\simeq& \frac{\widetilde{G}^2}{N}\sum_{a\neq0} \frac{\abs{\bra{\phi_0}\hat{J}\ket{\phi_a}}^2\abs{\bra{2}\hat{b}^{\dagger}\hat{b}^{\dagger}\ket{0}}^2}{2\widetilde{\Omega}(\widetilde{\Omega} + E_a-E_0)}.
\end{eqnarray}
Then, we can obtain Eq.~\eqref{eq:bdagbdag} in the main text using Eq.~\eqref{eq:cur_fluc} and the approximation, $E_a-E_0 \rightarrow \Delta E$.

\section{Numerical convergence} \label{Appendix:B}
\begin{figure}[t]
\begin{center}
\includegraphics[width=0.45\textwidth]{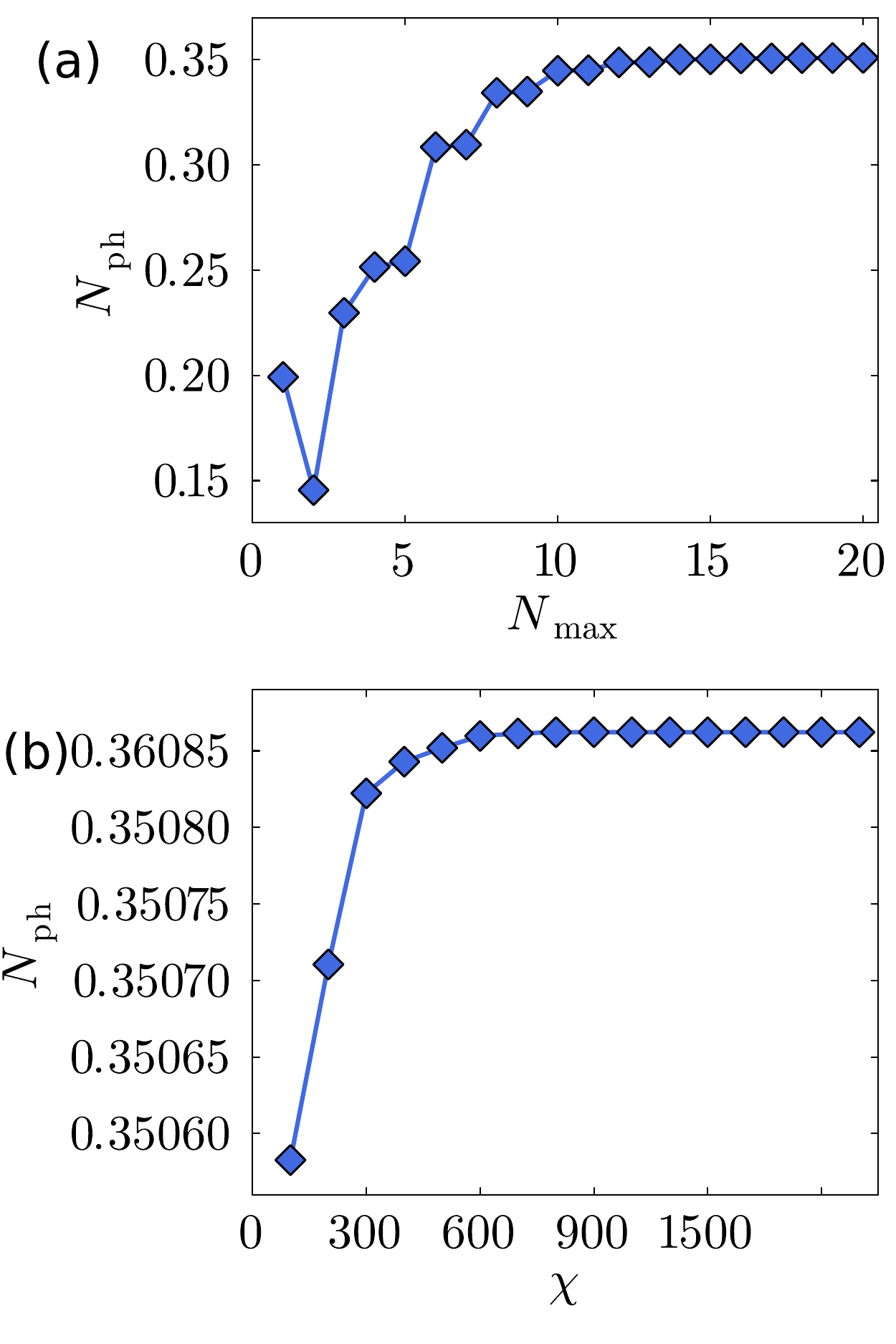}
\caption{
Photon number $N_{\mathrm{ph}}$ as a function of (a) the photon cutoff $N_{\mathrm{max}}$ and (b) the maximum bond dimension $\chi$ for the one-dimensional extended Hubbard model coupled to an optical cavity with $N=80$, $U/W=V/W=0$, $\Omega/W=10$, and $G=6$.
} \label{fig:convergence_DMRG}
\end{center}
\end{figure}
Here we confirm the convergence of the numerical calculations.
In the DMRG simulation, we have two parameters to control the accuracy of our numerical calculations, i.e., the maximum bond dimension $\chi$ and the photon number cutoff $N_{\mathrm{max}}$.
The maximum bond dimension $\chi$ determines the maximum dimension of the matrices in the MPS representation.
The photon cutoff determines the dimension of the local Hilbert space that describes the bosonic degree of freedom. 
Figure \ref{fig:convergence_DMRG} shows the $\chi$ and $N_{\mathrm{max}}$ dependence of the number of photons at $G=6$, which is the largest coupling constant considered in this study.
We show the result for $U/W=V/W=0$ because the number of photons is maximum at $G=6$ (Fig.~\hyperref[fig:Nph]{2(b)}).
We can see that the number of photons is converged at $\chi=1000$ and $N_{\mathrm{max}}=15$.

\begin{figure}[t]
\begin{center}
\includegraphics[width=0.45\textwidth]{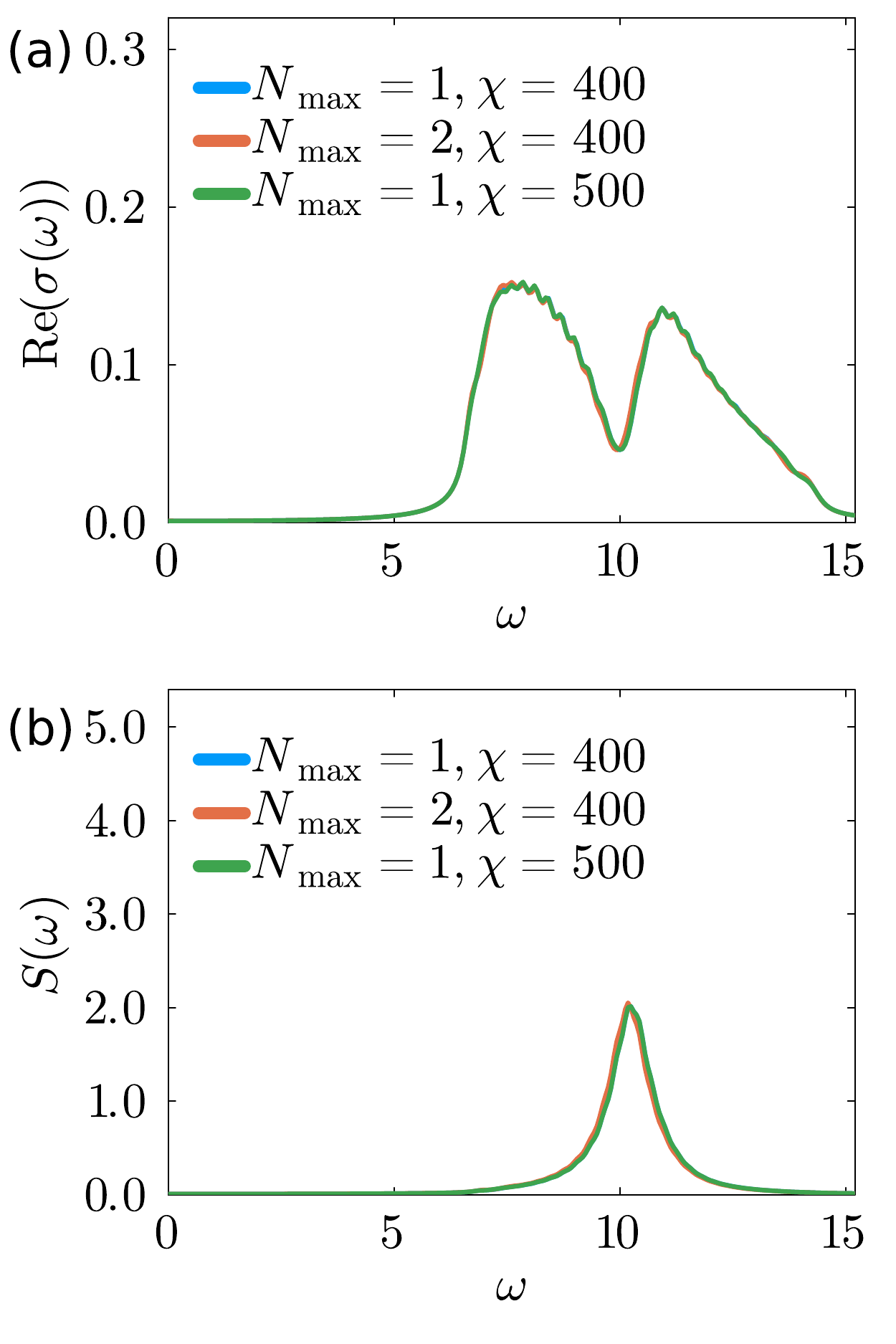}
\caption{
(a) Real part of the optical conductivity $\mathrm{Re}(\sigma(\omega))$ and (b) photon spectral functioin $S(\omega)$ for several $N_{\mathrm{max}}$ and $\chi$.
We set $N=40$, $U/W=10$, $V/W=0$, $\Omega/W=10$, and $G=0.5$
} \label{fig:convergence_TEBD}
\end{center}
\end{figure}
In the TEBD calculation, we also check the convergence in terms of $\chi$ and $N_{\mathrm{max}}$. 
During the time evolution, we fix the maximum bond dimension $\chi$. 
We set a relatively small coupling constant $G=0.5$ in the TEBD calculations, so that the necessary photon cutoff can be kept small.
Figure \ref{fig:convergence_TEBD} shows the $\chi$ and $N_{\mathrm{max}}$ dependence of the optical conductivity $\sigma(\omega)$ and photon spectral function $S(\omega)$ at $V=0$.
Due to high numerical cost of the TEBD simulation, we only perform the calculations up to $\chi=500$ and $N_{\mathrm{max}}=2$.
However, we can see that the $\sigma(\omega)$ and $S(\omega)$ are well converged at $\chi=400$ and $N_{\mathrm{max}}=2$.

\bibliography{reference}

\end{document}